\documentclass{article}

\usepackage{graphicx, import}
\usepackage{listings}
\usepackage[utf8]{inputenc}
\usepackage{float}
\usepackage{amsmath}
\usepackage{braket}
\graphicspath{{figures/}}
\usepackage{svg}
\usepackage[backend=biber, citestyle = numeric-comp]{biblatex}
\usepackage{caption}
\usepackage{hyperref}

\newcommand{\transP}{\mathcal{P}_{\mathit{fi}}}
\newcommand{\FtransP}{\mathcal{F}(\mathcal{P}_{\textit{fi}})} 

\newcommand{\wa}{\omega_s}
\newcommand{\wb}{\omega_i}
\newcommand{\wpump}{\omega_p}

\newcommand{\wzero}{\omega_0}

\newcommand{\wi}{\omega_i}
\newcommand{\ws}{\omega_s}
\newcommand{\intinf}{\int_{\rm -\infty}^\infty}	
\newcommand{\Te}{T_e}
\newcommand{\deltat}{\delta^{(t)}}
\newcommand{\Phiab}{\Phi(\wa,\wb)} 
\newcommand{\Dw}{\Delta_ \omega} 
\newcommand{\Dtau}{\Delta_ \tau} 
\newcommand{\wres}{\omega_{\rm res}} 
\newcommand{\Dj}{\Delta_j} 
\newcommand{\ej}{\epsilon_j} 
\newcommand{\ei}{\epsilon_i} 
\newcommand{\ef}{\epsilon_f} 

\DeclareMathOperator{\sinc}{sinc}

\DeclareMathOperator{\eV}{eV}
\DeclareMathOperator{\meV}{meV}

\DeclareMathOperator{\nm}{nm}
\DeclareMathOperator{\fs}{fs}

\begin{document}

\title{Entangled Two-Photon Absorption Spectroscopy with Varying Pump Wavelength}
\maketitle

\author{Lutz Mertenskötter\textsuperscript{1,2,*}, Kurt Busch\textsuperscript{1,3} and Roberto de J. León-Montiel\textsuperscript{4}}\\
\textit{\textsuperscript{1}Humboldt-Universität zu Berlin, Institut für Physik, AG Theoretische Optik \& Photonik, D-12489 Berlin, Germany\\
\textsuperscript{2}Weierstrass Institute for Applied Analysis and Stochastics, Mohrenstraße 39, 10117 Berlin
\textsuperscript{3}Max-Born-Institut, Max-Born-Stra{\ss}e 2A, 12489 Berlin, Germany\\
\textsuperscript{4}Instituto de Ciencias Nucleares, Universidad Nacional Autónoma de México, Apartado Postal 70-543, 04510 Cd. Mx., México}\\
\href{mailto:mertenskoetter@wias-berlin.de}{\textsuperscript{*}mertenskoetter@wias-berlin.de} 


\begin{abstract}
In virtual-state spectroscopy, information about the energy-level structure 
of an arbitrary sample is retrieved by Fourier transforming sets of measured 
two-photon absorption probabilities of entangled photon pairs where the degree 
of entanglement and the delay time between the photons have been varied. 
This works well for simple systems but quickly becomes rather difficult when 
many intermediate states are involved. We propose and discuss an extension 
of entangled two-photon absorption spectroscopy that solves this problem by 
means of repeated measurements at different pump wavelengths. Specifically, 
we demonstrate that our extension works well for a variety of realistic 
experimental setups. 
\end{abstract}


\section{Introduction}
Today, there exists a great variety of spectroscopic techniques, each with 
their own set of advantages and disadvantages, for a myriad of applications 
ranging from medicine 
\cite{luypaert2007near} 
and material science 
\cite{noda2005two} 
to biology 
\cite{tu1982raman} 
etc. 
Some of the more sophisticated protocols that have emerged are related to
two-photon spectroscopic techniques, where two timed photon pulses interact 
with the sample in short succession. 
Specifically, entangled two-photon absorption spectroscopy (eTPA spectroscopy) 
\cite{saleh,KOJIMA,nphoton,roberto_spectral_shape,dorfman2016,schlawin2017,oka2010,Schlawin-2018,villabona_calderon_2017,Varnavski2017,oka2018-1,oka2018-2,svozilik2018-1,svozilik2018-2,burdick2018,RobertoTemperatureControlled} 
represents a technique that utilizes the quantum nature of light to devise 
a powerful spectroscopic tool. For instance, it has been applied to propose novel experimental schemes that might be used for the 
determination of the electronic level structure of single molecules 
\cite{KOJIMA} 
and complex light-harvesting systems 
\cite{GoodsonTPAChromophores,paper2},
and has become a useful addition to the spectroscopic toolbox. 
In fact, eTPA spectroscopy as originally proposed by Saleh \emph{et al.} in 1998
\cite{saleh} 
relies on tuning and integrating over the entanglement time $T_e$ (a 
parameter of the second-order quantum correlation of the photon pair,
see section 2) in order to separate those features of the spectrum 
that allow direct access to the eigenenergies of the material from 
spurious background signals. 
However, this method can quickly become quite involved as it requires 
multiple experiments with two-photon states that bear different temporal 
correlations. Consequently, novel ways of extracting information from
eTPA signals have to be considered 
\cite{RobertoTemperatureControlled}. 

In this work, we develop a variant of this technique by exploiting 
the eTPA signal's dependence on the wavelength of the pump light.
More specifically, our proposed scheme extracts information about 
the electronic level structure of the samples under study by correlating 
measurements at two or more different pump wavelengths. Our setup 
could be realized using standard and widely used entangled-photon 
sources, thus opening up a novel avenue towards nonlinear quantum 
spectroscopy.

Our work is organized as follows. In section \ref{sec:the_model}, we describe the model
setup, the basic workings of ordinary eTPA spectroscopy, and elucidate
the problem of many intermediate states. 
We introduce our extension of eTPA spectroscopy to multiple pump 
wavelengths in section \ref{sec:Dependence of stuff on the pump wavelength} and provide a detailed discussion of its
applicability in realistic settings. Finally, we summarize our 
findings and conclude in section \ref{sec:conclusion}.

\section{The Model} \label{sec:the_model}
Our model setup consists of a source of entangled photon pairs with
tunable delay for two-photon absorption spectroscopy, a multi-level 
material system and a second-order perturbative analysis of the eTPA 
signals.

\begin{figure}[t!]
\centering
\includegraphics[width = \captionwidth]{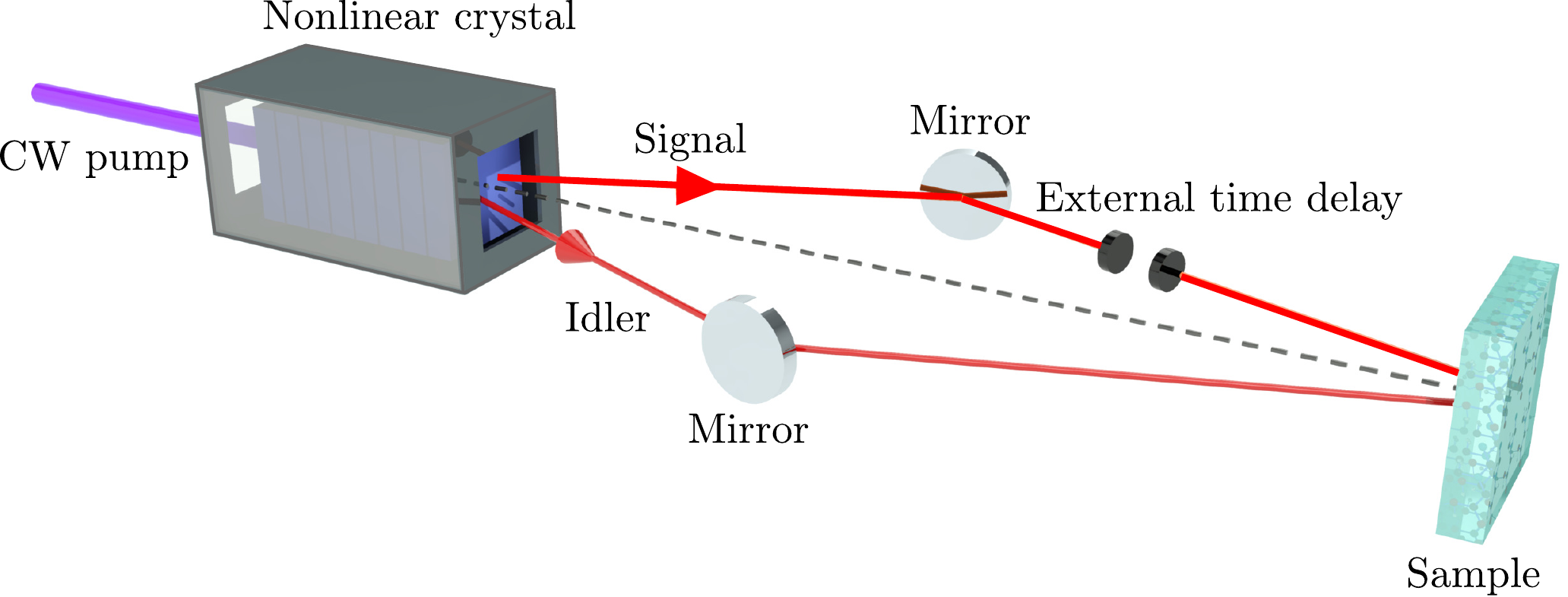}
\caption{Schematic of a eTPA spectroscopic setup: A collinear type-II SPDC 
         source is pumped  with monochromatic light of angular frequency 
				 $\wpump$ and produces two entangled photons with the frequencies 
				 $\ws$ and $\wi$ with a common central angular frequency 
				 $\wzero = \wpump/2$. 
				 A tunable delay $\tau$ is introduced into the path of one of 
				 the photons. Subsequently, both photons interact with a material 
				 system whose electronic level structure is schematically 
				 depicted in Fig. \ref{fig:onlyMatter}.}
\label{pathANDmatter}
\end{figure}
We consider an entangled-photon spectroscopy setup as schematically depicted 
in Fig. \ref{pathANDmatter}. The light source we employ is a two-photon state 
created by collinear Type-II spontaneous parametric down-conversion (SPDC) with continuous-wave pump and is described by the spectral decomposition 
\begin{align}
	\ket{\psi^F} = \intinf  \intinf d\wa d\wb \, 
	               \Phiab \, \hat a^\dagger_{\rm \wa} \hat a^\dagger_{\rm \wb} \ket 0,
\end{align} 
with $\hat a^\dagger_{\rm \wa}$, $\hat a^\dagger_{\rm \wb}$ being the creation operators for the signal ($s$) and idler ($i$) photons, respectively. The joint spectral function of the photons is given by
\begin{align}
	\Phi(\wa, \wb) &= \left(\frac{T_e}{\sqrt{\pi}}\right)^{1/2} 
	                  \delta(\wpump - (\wa + \wb)) 
										\sinc \left(\frac{T_e }{2}(\wa-\wb)\right) e^{i\wa \tau},
\end{align}
which is commonly referred to as the twin state 
\cite{salehTransparency}. 
Here, $l$ denotes the path length within the birefringent nonlinear crystal, 
$\wpump$ is  the angular frequency of the (monochromatic) light used to pump 
the SPDC source, $\omega_{\rm s}$ and $\omega_{\rm i}$ are the angular frequency 
of the signal and idler down-converted photons, respectively and $\tau$ is the external delay introduced into the path of the signal photon. 
Furthermore, the entanglement time $T_e$ is 
\begin{align}
	T_e = l(N_s-N_i)/2,
\end{align}
with the inverse group velocities $N_s = 1/v_{\rm g,s}$ ($N_i = 1/v_{\rm g,i}$) 
of the signal(idler) photons. \textcolor{black}{Note that, in the analysis below, we will assume that the photons are degenerate with central wave-packet frequencies $\ws^0,\wi^0 = \wzero = \wpump/2$.}\\

\begin{figure}[t!]
\centering
\includegraphics[scale=1]{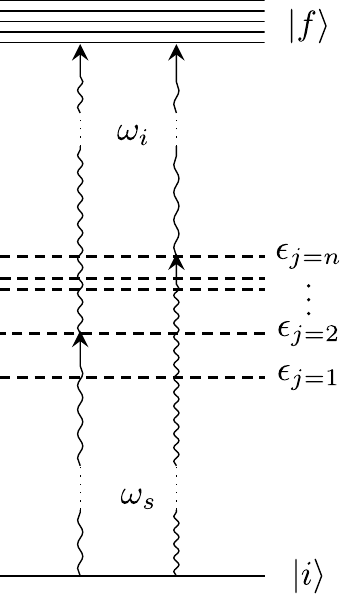}
\caption{Multi-level system including a number of intermediate states 
         with energies $\ej$ and a band of final states $\ket f$. 
				 The entangled photons are frequency-anti-correlated 
				 and satisfy $\ws +\wi = \epsilon_f - \epsilon_i$. 
				 The arrows indicate two possible transition pathways mediated by different intermediate states.}
\label{fig:onlyMatter}
\end{figure}

The sample material model is a multi-level system with non-degenerate 
energy eigenstates $\ket j$ with respective energies $\hbar \ej$ (see 
Fig.\ref{fig:onlyMatter}). It is described by the Hamiltonian
\begin{equation}
	\hat H_0 = \hat H_A = \sum_j \hbar \epsilon_j \ket j \bra j.
\end{equation}
Two of these states fulfill the two-photon resonance condition: 
\begin{align}
	\epsilon_f-\epsilon_i = 2\wzero = \wpump,
	\label{eq:twophotonresonance}
\end{align} 
and we consider them as the initial state $\ket i$ and the final state 
$\ket f$. The final state is assumed to lie within a band of closely 
spaced levels. It is important to note that in any realization of our 
setup $\ket f$ is defined by our choice of $\wpump$. The remaining $N$ 
states are intermediate states that contribute as pathways to the 
two-photon absorption signal, i.e the eTPA probability. 
These intermediate states are virtual states in the sense that they 
are energy eigenstates states $\ej$ of the unperturbed system, whose 
detuning to the center frequencies $\wzero$ of the entangled photons 
is larger than two times the Rabi frequency 
\cite{defVirtualStates}.

Using second-order perturbation theory, we can calculate the two-photon 
transition probabilities $\transP$ from the initial to the final state 
upon interaction with the twin field state.
Within dipole- and rotating-wave approximation, the perturbation, i.e. 
the interaction Hamiltonian, is described in the interaction picture as
\begin{align} 
	\hat V(t) = \hat  \mu(t) \hat E^{+}(t) 
	       = \left(\frac{\hbar \wzero}{4\pi\epsilon_0 cA}\right)^{1/2} 
				   \intinf d\omega  \, \hat \mu(t) \hat a_{\rm \omega} (t) + H.c.,\
\end{align}
where $\hat{\mu(t)}$ denotes the dipole operator.

Through a Fourier transform in conjunction with suitable coordinate 
transformations, the time-ordered time integral that arises in 
second-order perturbation theory evaluates to \cite{RobertoTemperatureControlled}
\begin{align} 
	\transP & =
            \left( \frac{\wpump ^2}{4\hbar^2\sqrt{\pi}\epsilon_0^2A^2T_e} \right)
					    \, 2\pi t\,\deltat(\ef - \ei-\wpump) \, s(T_e,\tau),
\end{align}
with the delta-like function of width $4/\pi t$
\begin{equation}
	\deltat(\ef - \ei -\wpump) = \frac{2\sin^2((\ef - \ei -\wpump) t/2)}{\pi t(\ef - \ei -\wpump)^2}
\end{equation}
and 
\begin{equation}
	s(T_e,\tau) = \left\vert \sum_{\rm j}  A_j (2 - e^{-i\Dj (T_e+\tau)} - e^{-i\Dj (T_e-\tau)}) \right\vert^2.
\end{equation}
In these expressions, we denote the energy mismatch of the center 
frequencies of the entangled photons $\wzero$, the intermediate 
states by $\Dj = \ej - \epsilon_i - \wzero$ and the transition 
matrix elements are $A_j = \mu_{\mathit{fj}} \mu_{\mathit{ji}}/\Dj$, where 
$\mu_{\mathit{kl}} = \bra k \hat{\mu} \ket l$ are the corresponding \textcolor{black}{transition} dipole 
moments. The delta-like function ensures energy conservation for 
times large compared to the energy mismatch of the pump angular 
frequency and the energy of the total transition.

Expanding the absorption cross section $s(T_e,\tau)$ now gives us
\begin{align}
	\begin{split}
		s(T_e,\tau) = \sum_{\rm j,k} A_j A_k^* \left( 4 - \right . & [e^{i\Dj-\Delta_k\tau} + c.c.] 
		                                          - 2\,e^{-i\Dj \Te} [e^{i\Dj\tau} + c.c.]  \\
                  - & \left.   2\,e^{\Delta_k \Te} [e^{i\Delta_k\tau} + c.c.] 
									           - e^{-(\Dj - \Delta_k \Te)} [e^{i\Dj+\Delta_k\tau} + c.c.] \right) \\
	\end{split}
\end{align}
Note, that for $N$ intermediate states $\ej$, the Fourier transform 
$\FtransP$ with respect to $\tau$, i.e. the eTPA spectrum, shows peaks 
at zero angular frequency, as well as the $2(N+1)N$ angular frequencies
\begin{align}
	\label{frequenciesPlain1}
	\pm \Dj =& \pm (\ej - \wzero),\\ 
	\label{frequenciesPlain2}
	\pm (\Dj - \Delta_k) =& \pm (\ej - \epsilon_k),\\
	\label{frequenciesPlain3}
	\pm (\Dj + \Delta_k) =&\pm (\ej + \epsilon_k - 2\wzero),
\end{align}
where $\ej$, $\epsilon_k$ \textcolor{black}{denote} the energies of the intermediate states.

\begin{figure}[H]
\centering

\includegraphics[width = \captionwidth]{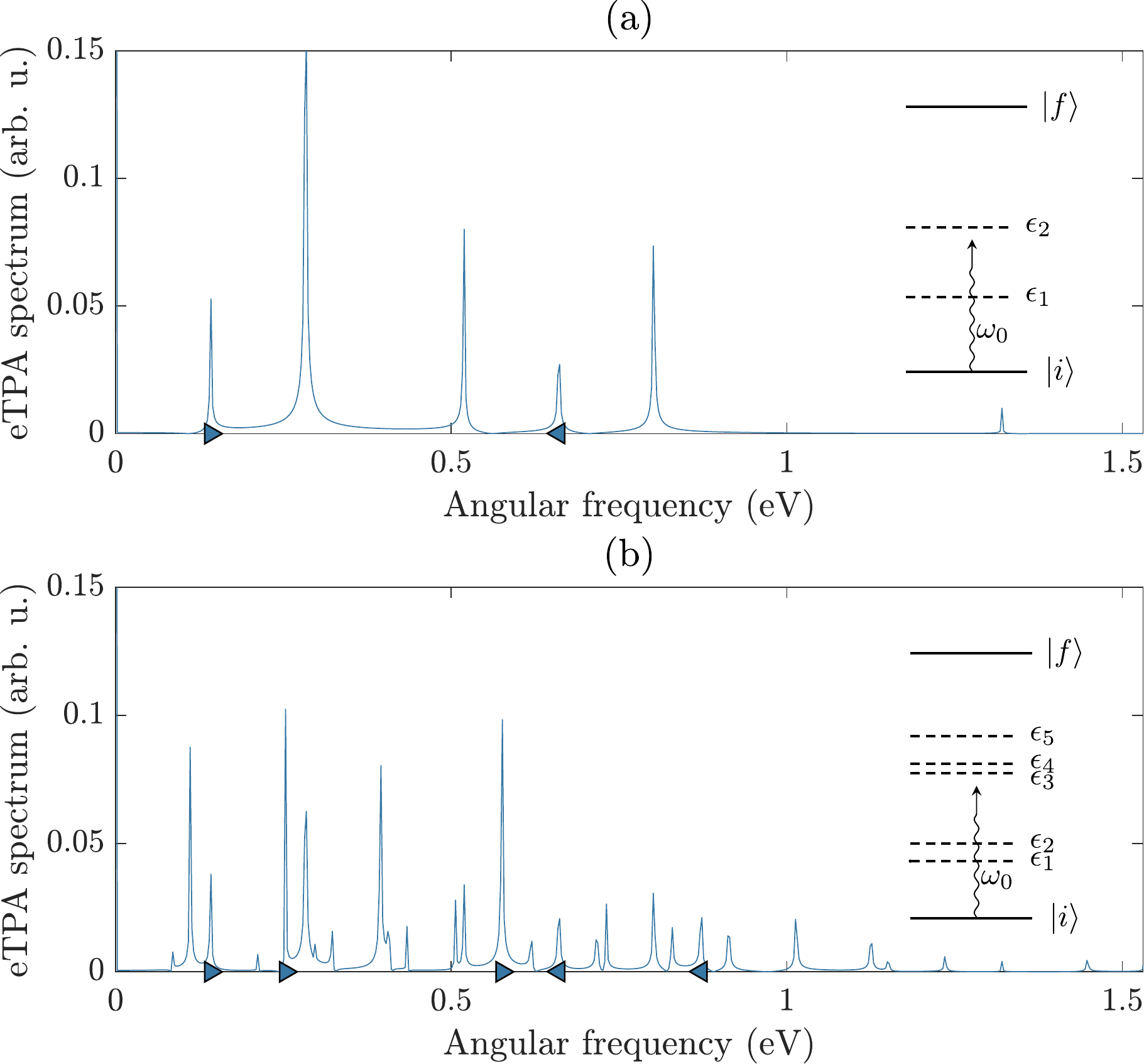}
\caption{Spectrum of the eTPA signal of two different matter systems. 
         a) two intermediate states at $\epsilon_1 = 0.86 \eV$ and 
				    $\epsilon_2 = 1.67 \eV$ and 
				 b) five intermediate states at $\epsilon_1 = 0.66 \eV$, 
				    $\epsilon_2 = 0.87 \eV$, $\epsilon_3 = 1.67 \eV$, 
					  $\epsilon_4 = 1.78 \eV$, and $ \epsilon_5 = 2.11 \eV$. 
				 The energy mismatches $+\Dj$ ($-\Dj$) are indicated by the right 
				 (left) facing triangles. Note that the number of peaks in the eTPA 
				 spectrum grows quadratically with the number of intermediate states. 
				 With a growing number of intermediate states it quickly becomes 
				 difficult to deduce the underlying energy levels from such spectra.}
\label{fig:fullExampleTwoStates}
\end{figure}

In order to illustrate how the different frequency peaks appear in the 
Fourier transform of the eTPA signal, we display in Fig. \ref{fig:fullExampleTwoStates}(a)
a system with two randomly-selected intermediate states for a fixed 
value of the pump wavelength of $\lambda_p = 405\nm$, corresponding 
to a central angular frequency of the entangled photons of 
$\wzero^1 = c\pi/\lambda_p = 1.53 \eV$. For simplicity, all dipole
moments have been set to the same constant value for all transitions.

We observe, that even though the energy mismatch $\Delta_1$ of the 
lower intermediate state is much larger than the energy mismatch 
$\Delta_2$ of the higher intermediate states, the corresponding peaks 
of the spectrum (marked with the triangles) do not differ 
proportionally in size. 
This suggests that, as a general rule, the heights of the peaks 
are not a reliable way of making sense of the spectrum and do not 
allow to deduce the underlying energy structure of the sample 
\cite{saleh,roberto_spectral_shape}.

In Fig. \ref{fig:fullExampleTwoStates}(b) we infer that it quickly
becomes difficult to interpret the spectrum when the number of 
intermediate states grows. This results from the fact that the number
of spectral peaks grows quadratically with the number of immediate
levels. Clearly, this severly limits the usefulness of the eTPA 
spectroscopy in the present form. 

\section{Extracting energies of intermediate-state levels}
\label{sec:Dependence of stuff on the pump wavelength}
Our goal is to extract the energies of the intermediate states $\ej$ 
from the eTPA spectrum. An easy way to achieve this, would be to identify 
within these peaks those at the frequencies $\Dj$ of Eq. (\ref{frequenciesPlain1}), 
as these only depend on one of the eigenenergies $\ej$ and its respective 
value is readily extracted by adding $\wzero$. For systems with a small 
number of intermediate states it may also be feasible to find the correct 
values by guessing the $N$ intermediate states as follows. First, we would
guess the $\Dj$ and then plug them into Eqs. (\ref{frequenciesPlain1})-(\ref{frequenciesPlain3}) 
and check whether the resulting frequencies line up with the actual 
spectrum. This approach can easily be carried out with simple spectra 
such as that of Fig. \ref{fig:fullExampleTwoStates}(a). 

However, in the general situation there are $2(N+1)N$ peaks and for the 
aforementioned technique of 'educated guessing', we would have to select 
$N$ members from this set and check whether they align with the actual
spectrum. Clearly, this scheme quickly becomes rather cumbersome to execute, 
see Fig. \ref{fig:fullExampleTwoStates}(b). Moreover, for systems with many 
intermediate states another detrimental effect sets in and fundamentally
obstructs the extraction of relevant information from a spectrum.
Specifically, as the number $N$ of peaks increases, it becomes more and 
more likely to encounter overlapping peaks with low amplitudes or very 
shallow signals that get lost in the background noise so that it will 
become less and less likely that the approach of 'educated guessing' will
succeed. The immediate response to this challenge would be to reduce the
noise floor and to increase spectral resolution but there clearly are
limits to what reasonably can be done. In what follows, we, therefore,
address this problem by extending eTPA spectroscopy by means of repeated
measurements at different pump wavelengths.

\subsection{Dependence of $s(T_e,\tau)$ on the pump wavelength}
\begin{figure}[t!]
	\centering
	\includegraphics[width = \captionwidth]{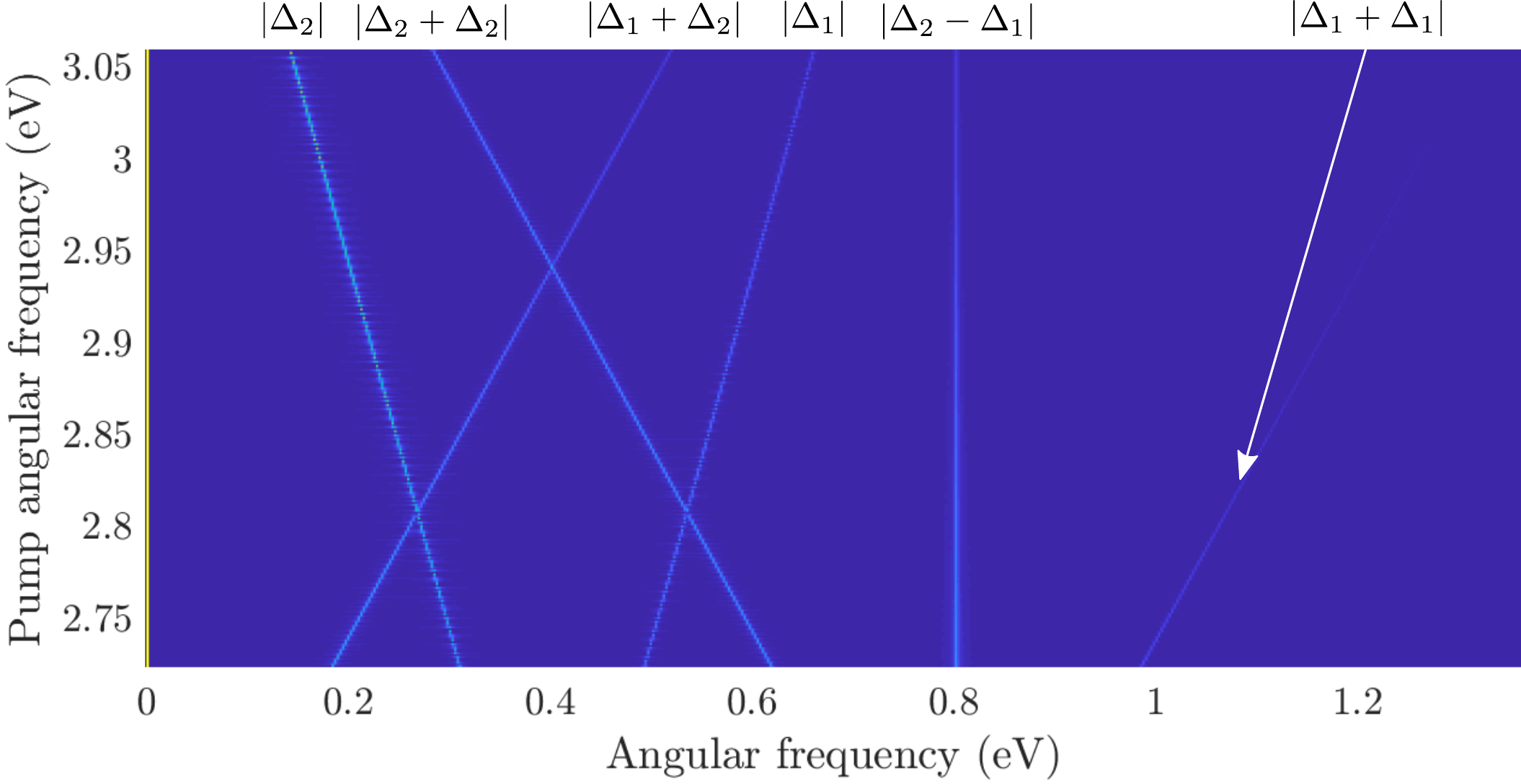}
	\caption{Spectrum of the eTPA signal at a range of $\wzero$ starting at 
					 $\wpump = 2\,\wzero^1$, i.e. the top edge of the plot corresponds 
				   to Fig. \ref{fig:fullExampleTwoStates}(a). 
				   At the top, the peaks are labeled in terms of the sets of 
				   frequencies in Eqs. (\ref{frequenciesPlain1})-(\ref{frequenciesPlain3}). 
				   We observe that the three distinct slopes clearly identify 
				   each peak as a member of one of the three distinct sets.}
	\label{peaklines}
\end{figure}
Fortunately, the sets of frequencies in Eqs. (\ref{frequenciesPlain1})-(\ref{frequenciesPlain3}) 
are set apart from all other frequencies by their dependence on 
$\wzero$ as demonstrated in Fig. \ref{peaklines}. Most importantly, 
the locations of the $+\Dj$ signals in Eq. (\ref{frequenciesPlain1}) 
go with $-\wzero$. This implies that by measuring at different pump 
wavelengths $\lambda_p$, i.e. different $\wzero$, we will be able 
to uniquely identify the intermediate-state energies of the sample, 
as we can distinguish them from the peaks at Eq. (\ref{frequenciesPlain2}), 
which do not depend on $\wzero$, and the peaks at Eq. (\ref{frequenciesPlain3}) 
which change with $2\wzero$.

\begin{figure}[H]
	\centering
	\includegraphics[width = \captionwidth]{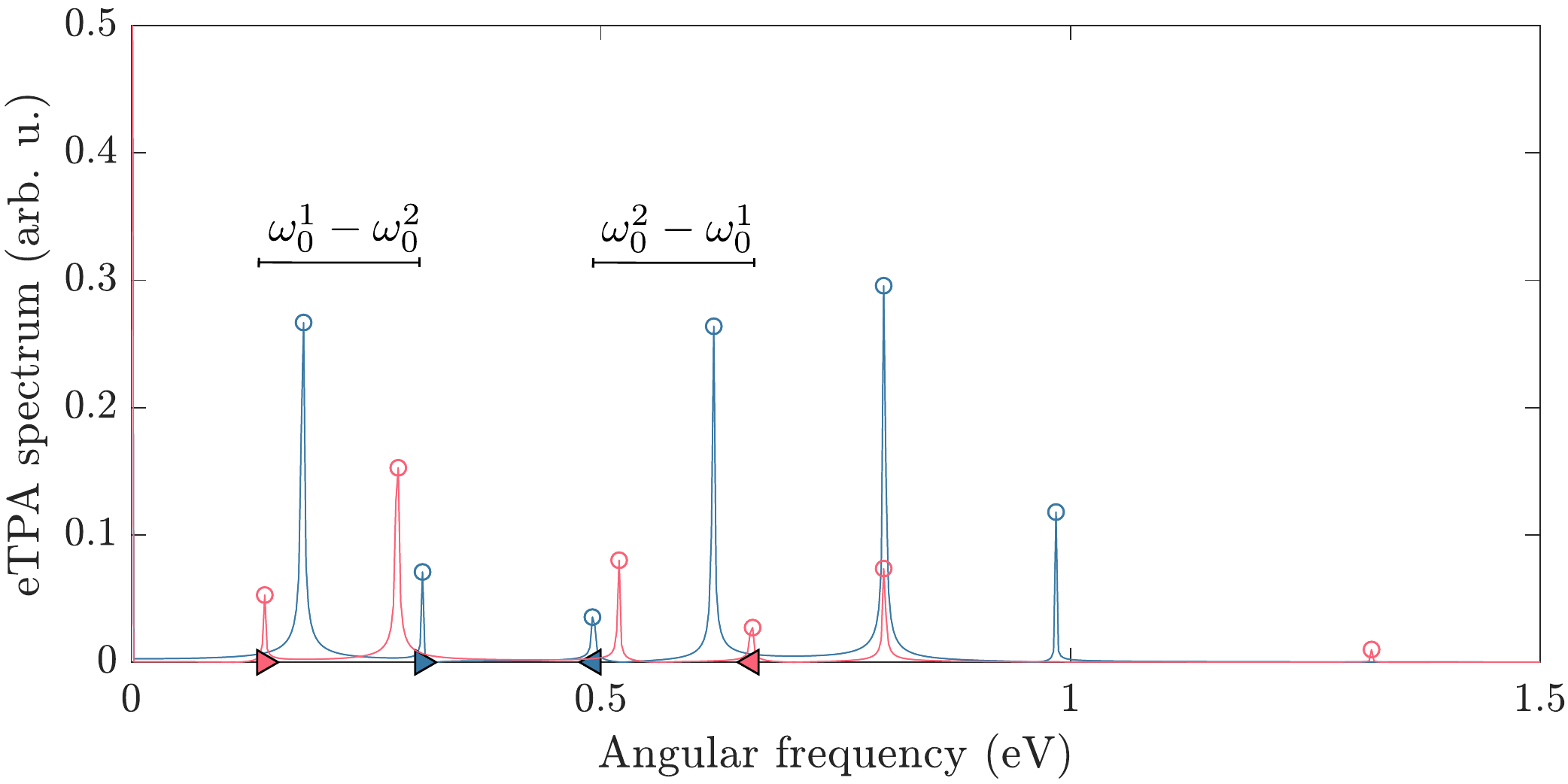}
	\caption{Spectrum of the eTPA signal at $\wzero^1 =1.53\eV$ (blue) 
					 and $\wzero^2 = 1.36\eV$  (red). The circles represent 
					 the peaks identified by our signal processing routine. 
					 The energy mismatches $+\Dj$ ($-\Dj$) are indicated by 
					 the right (left) facing triangles. 
					 Note, that exactly two pairs of peaks from the different 
					 spectra are separated by $\pm (\wzero^1 - \wzero^2)$, 
					 allowing us to identify the two intermediate-state 
					 energies $\ej$.}
	\label{setsEnough}
\end{figure}

In Fig. \ref{setsEnough} we display the eTPA spectrum for the same 
two-intermediate-state samples as in Fig. \ref{fig:fullExampleTwoStates}(a), 
considering two different central frequencies of the pump. Note, 
that in these plots the peaks corresponding to the frequencies $+\Dj$ 
of the two spectra are separated by a distance $\pm (\wzero^1 - \wzero^2)$. 
Now, we simply run a signal processing routine to identify the set of 
peaks of both spectra and find pairs, one element from each spectrum, 
that are separated by $\pm (\wzero^1 - \wzero^2)$. By adding the 
respective $\wzero$ to these peaks we can thus find the intermediate 
states $\ej = \Dj +\wzero$. \textcolor{black}{This process can easily be automated.}

An important advantage of this technique is that we can make further 
measurements at additional pump wavelengths should two measurements 
be insufficient to deduce the $\ej$ from the spectrum. 
This is preferable over simply increasing the resolution in the delay
time $\tau$, and decreasing statistical errors through repeated measurements 
of the same system, as features of the spectrum that are obscured by the 
overlapping of the peaks or low peak amplitudes at a particular frequency tend not to overlap or to be poorly visible at another pump frequency. This is due to the fact that peak positions and 
amplitudes also change with the pump frequency.
 
\subsection{Discrete Fourier transform and experimental accessibility of our technique}
\label{sec:Experimental_Accesibilty}
While the basic scheme laid out here is rather simple, a number of potential
problems lie in the choice of parameters for the experiment. In an actual
experiment, the values of $\tau$ are discrete. Assuming a free delay line 
with a mirror setup on a translation stage, their spacing $\Delta_r$ is 
determined by the smallest path delay we can introduce. Here we are using 
a value of $\Dtau = 0.3\times 10^{-15}$ s, which, using a mirror, translates 
to a step size $\Delta_L = c\Dtau/2 = 45\nm$, which is attainable using 
modern translation stages 
\cite{subNano}.

\begin{figure}[H]
	\centering
	\includegraphics[width =\captionwidth]{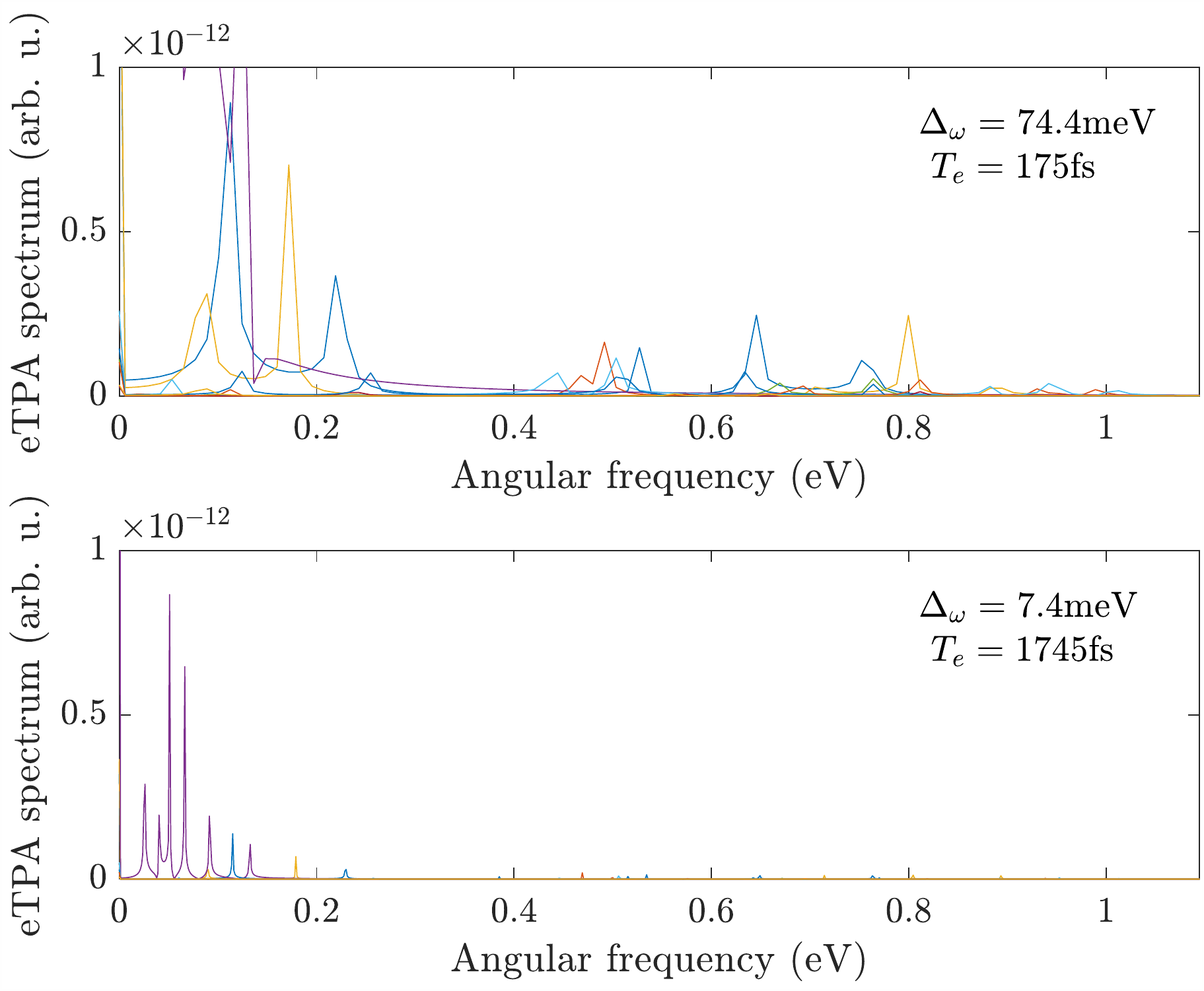}
	\caption{eTPA spectra for ten random sets of two intermediate states 
	         each for two different choices of $\Dw$ ($T_e$) at constant 
					 $\Delta_L = 45\nm$ and pump wavelength $\lambda_p = 405\nm$. 
					 We observe that a small bandwidth leads to strong attenuation 
					 of peaks at frequencies far from the resonance (bottom), yet 
					 a large bandwidth results in poor resolution in $\omega$ (top).}
	\label{compareDw}
\end{figure}

As we are using a discrete Fourier transform, our angular frequency resolution 
$\wres$ is defined by our sampling rate in time, i.e. smallest path delay, 
$\Dtau$ and the number of points we can measure $\tau_N =\tau_{\rm max}-\tau_{\rm min}/\Delta_{\rm \tau}$ 
by 
\begin{align}
	\wres  = \frac{\Dtau}{\tau_N} = \frac{1}{\tau_{\rm max}-\tau_{\rm min}} > \frac{\Delta_\omega}{2\pi}. 
	\label{wres<Dw}
\end{align} 
The inequality follows from the fact that the range of $\tau$ we can access 
is, in turn, limited by the bandwidth $\Delta_\omega$ of our entangled 
photons, as it defines their entanglement time $T_e$. Here, we assume an 
SPDC type-II source with a bandwidth of $\Delta_\omega = 7.4 \meV$
\cite{Fedrizzi_2009},  
resulting in an entanglement time of 
\begin{align}
	T_e = \frac{\pi}{\Dw} \approx 1745\fs.
\end{align}	
The two photons have to overlap in space-time to contribute to two-photon 
absorption and thus we have 
\begin{align}
	\tau_{\rm max} = -\tau_{\rm min} < T_e.
\end{align}
Furthermore, peaks that are supposed to be measurable with a simple setup 
need to roughly lie within the bandwidth $\Dw$ of our photons. This is a 
serious constraint, as at the same time our angular frequency resolution 
$\wres$ becomes poorer for large bandwidth and, consequently, small 
entanglement times $T_e$ [see Eq. (\ref{wres<Dw})]. 
In other words, ideally we would want a large $\Dw$ and a small $\wres$, 
which by (\ref{wres<Dw}) is mutually exclusive. In Fig. \ref{compareDw}, 
we illustrate this effect for two choices of $\Dw$.

This problem could be addressed by increasing the photon flux to offset 
the limited bandwidth and increase visibility of otherwise very low peaks. 
However, when choosing a bright source, we must take care to not exceed 
intensities $\Phi$ for which the quantum processes still cease to dominate
the absorption rate 
\cite{salehTransparency}.
Specifically, the absorption cross section $R$ has two contributions
\begin{align}
	R = R_e + R_r = \sigma_e \Phi +\delta_r \Phi^2, 
\end{align} 
where $\delta_r$ is the classical, i.e. probabilistic, absorption cross 
section and $\sigma_e \propto \transP$ is the quantum-mechanical cross 
section. It is the latter which we are trying to measure. Their actual
values depend on the experiment and have been analysed in detail in Refs. 
\cite{saleh, raymer2021entangled, landes2021quantifying}.

\textcolor{black}{Finally, it is worth remarking that} the angular frequency range of the eTPA spectrum is determined by our sampling rate in time as 
\begin{align}
	\omega_{\rm max} = \frac{2\pi}{2 \Dtau} = -\omega_{\rm min}
\end{align} 
and does not tend to be a limiting factor on our choice of parameters.

\section{Conclusion} \label{sec:conclusion}
We have demonstrated that the pump frequency $\wpump$ of a type-II SPDC 
source represents an additional resource for eTPA spectroscopy. Specifically,
we have shown that a varying pump wavelength provides a robust way to 
interpret the spectroscopic data that otherwise may well be very difficult
to interpret. In particular, for samples with complex energy spectra and 
when many intermediate states contribute to the two-photon absorption 
our novel approach can make eTPA spectroscopy feasible. 
Further, our analyses of the limitations in the choice of parameters 
have revealed that there is ample room for balanced choices regarding 
frequency resolution as well as the frequency range. How these are 
weighted depends on the concrete problem at hand. Further, the trade-off 
between resolution and range can, to some extent, be relaxed by reducing 
the step size $\Delta_L$ using more sophisticated delay lines. 

\section{Acknowledgments}
We would like to thank Armando Perez-Leija for many fruitful discussions 
on the topic, as well Sven Ramelow for his valuable insights on the 
experimental side of things.
\section{Funding}
 R.J.L.-M. thankfully acknowledges financial support by CONACyT under the project CB-2016-01/284372 and by DGAPA-UNAM under the project UNAM-PAPIIT IN102920; 
 L. M. acknowledges funding by the Deutsche Forschungsgemeinschaft (DFG, German Research Foundation) under Germany's Excellence Strategy – The Berlin Mathematics Research Center MATH+ (EXC-2046/1, project ID: 390685689); K.B. and R.J.L.-M. acknowledge financial support by the
 Deutsche Forschungsgemeinschaft (DFG), SFB 951, (Project 182087777).

\printbibliography
\end{document}